\documentclass{ws-p8-50x6-00}

\begin{document}

\title{Particle Production in High-energy Heavy-ion Collisions}

\author{Xin-Nian Wang}

\address{Nuclear Science Division, Lawrence Berkeley National Laboratory,
Berkeley, California, 94720 \\ 
Department of Physics, Shandong University, Jinan, P.R.China, 250100}


\maketitle

\abstracts{Particle production mechanisms in high-energy heavy-ion
collisions are reviewed in connection with recent experimental data
from RHIC. Implications on mini-jet production, parton saturation
and jet quenching are discussed.}

\section{Introduction}

Quark-gluon plasma (QGP) is expected to form in high-energy heavy-ion
collisions. Though some proposed signals can provide more direct 
measurements of the initial parton density, many must reply on
information inferred indirectly from the measurement of final hadron 
spectra. Therefore, both global observables such as the rapidity density 
of hadron multiplicity and high $p_T$ hadron spectra can provide
important links of a puzzle that can eventually lead one to a
more complete picture of the dynamics of heavy-ion collisions 
and formation of QGP.

Mini-jet production in a two-component model has long been proposed 
to explain the energy dependence of total cross section 
\cite{gaisser,pancheri} and particle production \cite{chen,wang91}
in high-energy hadron collisions. It has also been 
proposed \cite{eskola,bm} and incorporated 
in the HIJING model \cite{hijing,wang97} to describe initial 
parton production in high-energy heavy-ion collisions. In this
model, parton production from semi-hard processes becomes quite
significant, giving rise to the high initial parton density
in high-energy heavy-ion collisions. However, there are still
quite large uncertainties due to the lack of knowledge of the
initial parton distributions in nuclei and the final state
interaction of produced partons. Therefore, the study of energy 
and centrality dependence of central rapidity 
density \cite{wg01} can provide 
important constraints on models of initial entropy production and
shed lights on the initial parton distributions in nuclei.
For example, the available RHIC experimental data 
\cite{phob1,phob2,phenix,star,brahms} can already rule out the
simple two-component model without nuclear modification of the
parton distributions in nuclei \cite{wg01}. However, further
studies are needed to constrain the unknown nuclear shadowing 
of the gluon distribution in nuclei and to further distinguish
the conventional parton production mechanism from
other novel physics such as parton saturation \cite{kn,kl}. 

Among the produced particle, a few have very high transverse
momentum. They are related to high-energy jet production which is a very 
useful tool to study the properties of dense matter formed in 
high-energy heavy-ion collisions. Because
large $p_T$ partons are produced very early in heavy-ion collisions
and their production rates can be calibrated in $pp$ and $pA$ collisions
at the same energy, they are ideal probes of the dense matter that
is formed in the same reaction. What probes the dense medium is the
scattering induced energy loss suffered by an energetic parton as it 
propagates through the matter. The parton energy loss is directly
related to the parton density of the medium. These are the main topics
of this talk.

\section{Two-component model}

\begin{figure}[t]
%\figurebox{20pc}{15pc}{} 
% to have a box alone
\epsfxsize=15pc 
% will enlarge or reduce the postscript figures based on the xsize
\centerline{\epsfbox{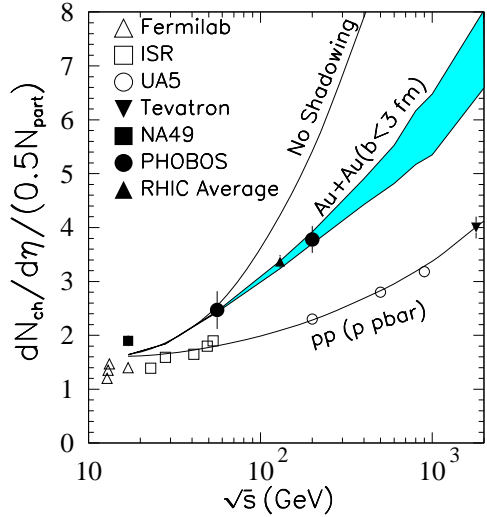}} % postscript image file name
\caption{
Charged particle rapidity density {\em per  participating 
nucleon pair} versus the c.m. energy. The RHIC
data\protect\cite{phob1,phob2} (filled circle and up-triangle) 
for the  6\% most central Au+Au are  
compared to $pp$ and $p\bar{p}$ data (open symbols)  
\protect\cite{ppbar,ppisr,ppfermilab} 
and the NA49 $Pb+Pb$(central 5\%) data \protect\cite{na49} (filled square).  
The two-component mini-jet model with and without shadowing is
also shown. The shaded area for central $Au+Au$ collisions corresponds
to the range of gluon shadowing parameter 
$s_g=0.24$--0.28 [Eq.~(\protect\ref{eq:shq})]. 
\label{fig1} }
\end{figure}

In the two-component model, the soft and hard processes are 
separated by a cut-off scale $p_0$. While the cross section of soft 
interaction $\sigma_{\rm soft}$ is considered nonperturbative and 
thus noncalculable, the jet production
cross section $\sigma_{\rm jet}$ is assumed to be given by 
perturbative QCD (pQCD) for transverse momentum transfer $p_T>p_0$.
The two parameters, $\sigma_{\rm soft}$ and $p_0$, are determined
phenomenologically by fitting the experimental data of total 
$p+p(\bar{p})$ cross sections within the two-component
model \cite{gaisser,pancheri,chen,wang91,hijing,wang97}.
The cut-off scale $p_0$, separating nonperturbative and pQCD
components, could in principle depend on both energy
and nuclear size. Using the Gluck-Reya-Vogt (GRV) 
parameterization \cite{grv} of parton distributions and 
following the same procedure as in the original HIJING \cite{hijing}, 
one finds that the cut-off scale must have a weak energy dependence,
\begin{equation}
p_0(\sqrt{s})=3.91-3.34\log(\log\sqrt{s})+0.98\log^2(\log\sqrt{s})
\nonumber \\
+0.23\log^3(\log\sqrt{s}),
\end{equation}
in order to fit the experimental data, because of the rapid 
increase of gluon distribution at small $x$.
Shown in Fig.~\ref{fig1} is the calculated central rapidity density,
\begin{equation}
\frac{dN_{\rm ch}}{d\eta}=\langle n\rangle_{\rm s}
+\langle n\rangle_{\rm h} 
\frac{\sigma_{\rm jet}(s)}{\sigma_{\rm in}(s)},
\end{equation}
for $p+p(\bar{p})$ collisions as a function of energy $\sqrt{s}$,
where $\langle n\rangle_{\rm s}=1.6$ and $\langle n\rangle_{\rm h}=2.2$
represent particle production from soft interaction and jet
hadronization, respectively. The jet cross section is the lowest order
pQCD result with a $K$-factor of 2 and GRV parton distributions.

To extrapolate the two-component model to nuclear collisions,
one assumes that multiple mini-jet production is incoherent and
thus is proportional to the number of binary collisions 
$N_{\rm binary}$. The soft interaction is however coherent and 
proportional to the number of participant nucleons $N_{\rm part}$
according to the wounded nucleon model \cite{bialas}. 
One important nuclear effect we have to consider in the two-component 
model is the nuclear shadowing of parton distributions in nuclei 
at small $x$. Such a nuclear shadowing effect in jet production 
can be taken into account by assuming modified parton distributions 
in nuclei,
\begin{equation}
f_a^A(x,Q^2)=AR_a^A(x,Q^2)f_a^N(x,Q^2).
\end{equation}
Using the experimental data from DIS off nuclear targets and unmodified
DGLAP evolution equations, one can
parameterize $R_a^A(x,Q^2)$ for different partons and nuclei.
Recent new data however indicate that the simple
parameterization for nuclear shadowing used in HIJING
is too strong \cite{liwang}
for heavy nuclei. Instead one has to use the following
new parameterization,
\begin{eqnarray}
R_a^A(x)&=&1.0+1.19\log^{1/6}\!\!\!A\;(x^3-1.2x^2+0.21x) \nonumber \\
        &-&s_a\;(A^{1/3}-1)^{0.6}(1-3.5\sqrt{x})\exp(-x^2/0.01)
\label{eq:shq}
\end{eqnarray}
with $s_q=0.1$ for all quark distributions and $s_g=0.24$--0.28 
for gluon. Assuming no
final state effects on multiplicity from jet hadronization, the 
rapidity density of hadron multiplicity in heavy-ion collisions 
as shown in Fig.~\ref{fig1} is,
\begin{equation}
\frac{dN_{ch}}{d\eta}=\frac{1}{2}\langle N_{\rm part}\rangle 
\langle n\rangle_{s} 
+ \langle n\rangle_{h}\langle N_{\rm binary}\rangle 
\frac{\sigma_{\rm jet}^{AA}(s)}{\sigma_{\rm in}},
\label{eq:nch} 
\end{equation}
where $\sigma_{\rm jet}^{AA}(s)$ is the averaged inclusive jet cross
section per $NN$ in $AA$ collisions including parton shadowing effect. 
The average number of participant nucleons and number of binary 
collisions for given impact-parameters can be estimated 
using HIJING Monte Carlo simulation.
It is clear that one has to consider parton shadowing in jet 
production in heavy-ion collisions.

\section{Parton Saturation}

Similar results of multiplicity in heavy-ion collisions are also 
predicted by other models \cite{ekrt,other},
in particular the initial-state parton saturation model \cite{kn,kl}. 
It is based on the nonlinear Yang-Mills field dynamics \cite{bm,mv}
assuming that nonlinear gluon interaction below a saturation scale
$Q_s^2\sim \alpha_s\, xG_A(x,Q_s^2)/\pi R_A^2$ leads to a classical
behavior of the gluonic field inside a large nucleus,
where $G_A(x,Q_s^2)$ is the gluon distribution at $x=2Q_s/\sqrt{s}$.
Assuming particle production in high-energy heavy-ion collisions 
is dominated by gluon production from the classical gluon field,
one has a simple form \cite{kl} for the charged hadron rapidity
density at $\eta=0$,
\begin{equation}
\frac{2}{\langle N_{\rm part}\rangle}\frac{dN_{ch}}{d\eta}=c
\left (\frac{s}{s_0}\right)^{\lambda/2}\left[
\log\left(\frac{Q_{0s}^2}{\Lambda_{\rm QCD}^2}\right)
+ \frac{\lambda}{2}\log\left(\frac{s}{s_0}\right) \right ],
\end{equation}
with $c\approx 0.82$ \cite{kn}. This is shown in Fig.~\ref{fig4}
as solid lines as compared to the two-component model (shaded area). 
Here, $\Lambda_{\rm QCD}=0.2$ GeV, $\lambda=0.25$ and 
the centrality dependence of the saturation scale $Q_{0s}^2$ at
$\sqrt{s_0}=130$ GeV is taken from Ref.~\cite{kn}. The two-component
model has an impact-parameter dependent parton shadowing.

\begin{figure}[t]
%\figurebox{20pc}{15pc}{} 
% to have a box alone
\epsfxsize=15pc 
% will enlarge or reduce the postscript figures based on the xsize
\centerline{\epsfbox{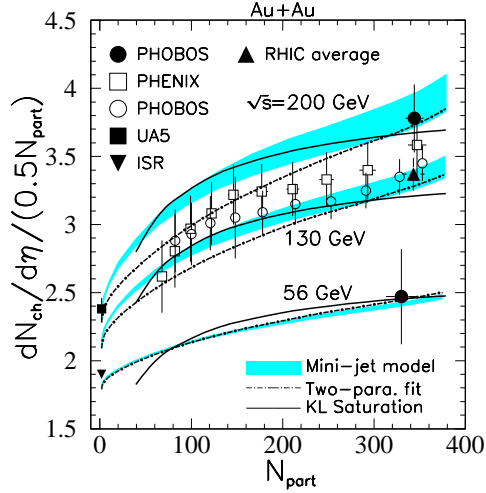}} % postscript image file name
\caption{
The charged hadron central rapidity density per participant nucleon 
pair as a function of the averaged number of participants from
the two-component model (shaded lines), two-parameter 
fit \protect\cite{liwang} (dot-dashed lines) and parton
saturation model \protect\cite{kl} as compared to experimental
data \protect\cite{phob1,phenix,ppbar,ppisr}.
\label{fig2} }
\end{figure}

Comparing the two model results in Fig.~\ref{fig2}, one notices that 
the saturation and two-component model agree with each other
in most regions of centrality except very peripheral and very central
collisions. In central collisions, results of saturation model tend
to be flatter than the two-component model. In this region, there are
still strong fluctuations in parton production in the two-component
model through the fluctuation of $N_{\rm binary}$ while $N_{\rm part}$
is limited by its maximum value of $2A$. That is why
$dN_{ch}/d\eta/\langle N_{\rm part}\rangle$ continues to increase
with $\langle N_{\rm part}\rangle$ in the central region. Such a
fluctuation is not currently taken into account in the saturation
model calculation. More accurate measurements
with small errors (less than 5\%) will help to distinguish these two
different behaviors. For peripheral collisions, saturation model results
fall off more rapidly than the mini-jet results. However, the experimental
errors are very big in this region because of large uncertainties 
related to the determination of the number of participants. Therefore,
it will be very useful to have light-ion collisions at the same energy
to map out the nuclear dependence of the hadron multiplicity in this region.
An alternative is to study the ratios of hadron multiplicity of heavy-ion
collisions at two different energies as a function of 
centrality \cite{liwang}. In 
this case, the errors associated with the determination of centrality
will mostly cancel.

It is interesting to point out that in the saturation
model that assumes a particle production mechanism dominated by coherent 
mini-jet production below the saturation scale $Q_s$, the value
of $Q_s$ determined in Ref.\cite{kn,kl} is much smaller than the 
cut-off $p_0$ in the two-component model constrained by the $p+p(\bar{p})$
data. As demonstrated in this paper, the number of mini-jet production
below such scale is still very large and should contribute to the final
hadron multiplicity.

\section{Parton energy loss}

Among the bulk of produced hadrons in heavy-ion collisions, there are
also high $p_T$ hadrons produced in events with hard processes.
In principle, jet quenching effect
could also lead to increased total hadron multiplicity \cite{wg92}
due to the soft gluons from the bremsstrahlung. However, a recent
study \cite{ww01} of parton energy loss in a thermal environment 
found that the effective energy loss is significantly reduced for 
less energetic partons due to detailed balance by thermal absorption.
Thus, only large energy jets lose significant energy via gluon 
bremsstrahlung. Since the production rates of these large energy
jets are very small at the RHIC energy, their contributions to
the total hadron multiplicity via jet quenching should also be small.
Similarly we also assume that parton thermalization during the
early stage contributes little to the final hadron multiplicity.

Theoretical studies of the parton energy loss in hot medium date back to 
the first attempt by Bjorken \cite{bj} to calculate elastic energy
loss of a parton via elastic scattering in the hot medium. A simple
estimate can be given by the thermal averaged energy transfer
$\nu_{\rm el}\approx q_\perp^2/2\omega$ of the jet parton to a thermal
parton with energy $\omega$, $q_\perp$ being the transverse momentum
transfer of the elastic scattering. The resultant elastic energy 
loss \cite{wang97}
\begin{equation}
\frac{dE_{\rm el}}{dx}=C_2\frac{3\pi\alpha_{\rm s}^2}{2}T^2
\ln\left(\frac{3ET}{2\mu^2}\right)
\end{equation}
is sensitive to the temperature of the thermal medium but is in general 
small compared to radiative energy loss. Here, $\mu$ is the Debye screening
mass and $C_2$ is the Casimir of the propagating parton in its fundamental 
presentation. The elastic energy loss can also be calculated within finite
temperature QCD \cite{thoma} with a similar result, but with a more careful
and consistent treatment of screening effect.

Though there had been estimates of
the radiative parton energy loss using the uncertainty principle \cite{bh},
a first theoretical study of QCD radiative parton energy loss incorporating
Landau-Pomeranchuk-Migdal interference effect \cite{lpm} is by Gyulassy and
myself \cite{gw} where multiple parton scattering is modeled by a screened
Coulomb potential model. Baier {\it et al.} (BDMPS) \cite{bdmps} later 
considered the effect of gluon rescattering which turned out to be very 
important for gluon radiation induced by multiple scattering in a dense
medium. These two studies have ushered in many recent works on the subject,
including a path integral approach to the problem \cite{zakharov} and opacity 
expansion framework \cite{glv,wied} which is more suitable for multiple
parton scattering in a thin plasma. The radiative parton energy loss to the
leading order of the opacity $\bar{n}=L/\lambda$ in the thin plasma of size
$L$ is estimated as \cite{glv,ww01}
\begin{equation}
\frac{dE_{\rm rad}}{dx}\approx C_2 \frac{\alpha_{\rm s} \mu^2}{4}
\frac{L}{\lambda} \ln\left(\frac{2E}{\mu^2 L}\right),
\end{equation}
where $\lambda$ is the gluon's mean-free-path in the medium. The unique
$L$ dependence of the parton energy loss is a consequence of the non-Abelian
LMP interference effect in a QCD medium. It is also shown in a recent
study \cite{ww01} that thermal absorption and stimulated emission in a thermal
environment can be neglected for high energy partons ($E\gg \mu$) while
they are important for intermediate energy partons.

Using this latest result one can estimate the total energy loss for 
a parton with initial energy $E=40$ GeV to be about $\Delta E\approx 10$ GeV
after it propagates a distance of $L=6$ fm in a medium with $\mu=0.5$ GeV
and $\lambda=1$ fm. For an expanding system, the total energy loss
is reduced by a factor of $2\tau_0/L$ from the static 
value \cite{baier98,gvw}. Assuming that most of
this energy loss is carried by gluons outside the jet cone \cite{baier99},
measuring the energy loss would require the experimental resolution
$\delta E$ to be much smaller than the total energy loss $\Delta E$.
With the measured total multiplicity density $dN/d\eta \approx 900$
\cite{phob1} and energy density $dE_T/d\eta\approx 500$ GeV \cite{phet}
in central $Au+Au$ collisions at $\sqrt{s}=130$ GeV, one can estimate that
the average total background energy within the jet cone ($\delta\eta=1$
and $\delta\phi=1$) is about $\Sigma E_T\approx 80$ GeV with
a fluctuation of $\delta E_T\approx 10$ GeV. It is therefore very
difficult, if not impossible, to determine the energy of a jet on
a event-by-event base \cite{Wang:1990bk}. Since high $p_T$ hadrons
in hadron and nuclear collisions come from fragmentation of high $E_T$
jets, energy loss naturally leads to suppression of high $p_T$ hadron
spectra. Miklos Gyulassy and I then proposed \cite{wg92} that one has 
to reply on measuring the suppression of high $p_T$ hadrons to study 
parton energy loss in heavy-ion collisions. Since inclusive hadron
spectra is a convolution of jet production cross section and the
jet fragmentation function in pQCD, the suppression of inclusive high $p_T$
hadron spectra is a direct consequence of the medium modification of the
jet fragmentation function induced by parton energy loss.
Assuming that jet fragmentation function is the same for the final
leading parton with a reduced energy, the modified fragmentation 
function can be assumed as \cite{wh}
\begin{equation}
\widetilde{D}(z)\approx \frac{1}{1-\Delta E/E} 
D\left(\frac{z}{1-\Delta E/E}\right). \label{dbar0}
\end{equation}
Therefore, in this effective model the measured modification of fragmentation 
function can be directly related to the parton energy loss.

\section{Modified fragmentation functions}

Since a jet parton is always produced via a hard process involving a
large momentum scale, it should also have final state radiation with
and without rescattering leading to the DGLAP evolution equation of
fragmentation functions. Such final state radiation effectively
acts as a self-quenching mechanism softening the leading parton
momentum distribution. This process is quite similar to the induced 
gluon radiation and the two should have strong interference
effect \cite{glv,wied}. It is therefore natural to study jet quenching
and modified fragmentation function in the framework of modified
DGLAP evolution equations in a medium \cite{wgdis}.

The simplest case for jet quenching is deeply inelastic scattering
of an electron off a nucleus target where the virtual photon knocks
one quark out of a nucleon inside a nucleus. The quark then will have
to propagate through the rest of the nucleus and possibly scatter
again with other nucleons with induced gluon radiation. The
induced gluon radiation reduces the quark's energy before it fragments 
into hadrons with a modified fragmentation function. One can
study the nuclear modification of the fragmentation function by
comparing it with the same measurement in DIS with a nucleon target.

In an infinite momentum frame, where the photon carries momentum
$q=[-x_Bp^+,q^-,\vec{0}_\perp]$ and the momentum of the target
per nucleon is $p=[p^+,0,\vec{0}_\perp]$ with the Bjorken variable
defined as $x_B=Q^2/2q^-p^+$, one can calculate the modified
fragmentation function. Because of the LPM interference effect,
the formation time of the gluon radiation due to the 
LPM interference requires the radiated gluon to have a minimum transverse 
momentum $\ell_T^2\sim Q^2/MR_A\sim Q^2/A^{1/3}$. 
The nuclear corrections to the fragmentation function due to double 
parton scattering will then be in the order of 
$\alpha_s A^{1/3}/\ell_T^2 \sim \alpha_s A^{2/3}/Q^2$, which depends
quadratically on the nuclear size. For large values of
$A$ and $Q^2$, these corrections are leading and yet the requirement
$\ell_T^2\ll Q^2$ for the logarithmic approximation in deriving the 
modified fragmentation function is still valid.

\begin{figure}[t]
%\figurebox{20pc}{15pc}{} 
% to have a box alone
\epsfxsize=15pc 
% will enlarge or reduce the postscript figures based on the xsize
\centerline{\epsfbox{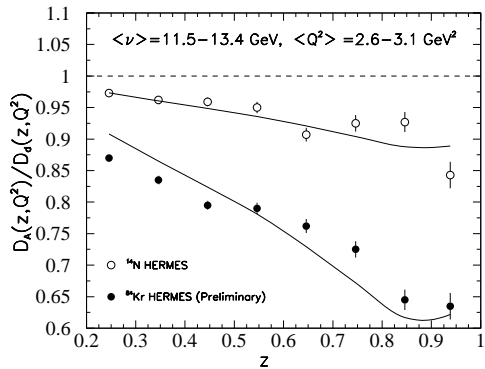}} % postscript image file name
\caption{
The predicted nuclear modification of jet fragmentation function
is compared to the HERMES data \protect\cite{hermes}.
\label{fig3}}
\end{figure}

Shown in Fig.~\ref{fig3} are the calculated nuclear modification 
factor of the fragmentation function for $^{14}N$ and $^{84}Kr$ 
targets as compared to the recent HERMES data \cite{hermes}. 
The predicted shape of the $z$ dependence 
and the quadratic nuclear size dependence agrees well with the 
experimental data. The energy 
dependence of the suppression also has excellent agreement with our 
prediction. What is amazing is the clear 
quadratic $A^{2/3}$ nuclear size dependence of the suppression
which is a true QCD non-Abelian effect. 
In fitting the data of the overall 
suppression for $^{14}N$ target we obtain the only parameter in 
our calculation, $\widetilde{C}\alpha^2_{\rm s}=0.00065$ GeV$^2$.

One can also calculate theoretically the average energy loss by 
the quark, which is the energy carried away by the radiated gluons,
 \begin{eqnarray}
\Delta E=\nu\langle\Delta z_g\rangle
\approx  \widetilde{C} (C_A\alpha_{\rm s}^2/N_c) 3\ln(1/2x_B).
\end{eqnarray}
With the value of $\alpha_{\rm s}^2\widetilde{C}$, one gets the
quark energy loss $dE/dx\approx 0.3$ GeV/fm for the $Kr$ target.

\section{Jet quenching in heavy-ion collisions}

In high-energy heavy-ion collisions, the jet production rate is
not affected by the formation of dense matter and the final state
multiple scattering. One can assume that the high $p_T$ hadron
spectra can then be given by the convolution of the jet production
cross section and the medium modified jet fragmentation 
function $\widetilde{D}_{h/c}(z_c,Q^2)$.

\begin{figure}[t]
%\figurebox{20pc}{15pc}{} 
% to have a box alone
\epsfxsize=15pc 
% will enlarge or reduce the postscript figures based on the xsize
\centerline{\epsfbox{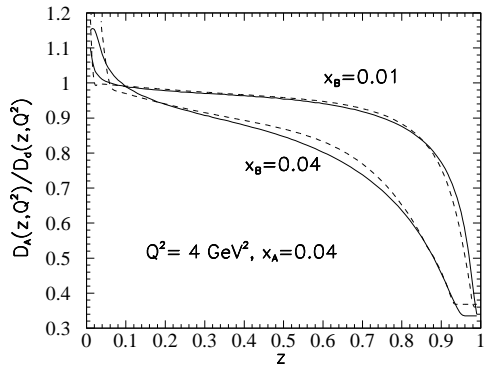}} % postscript image file name
\caption{
Comparison of the calculated nuclear modification with the
effective model in Eq.~(\ref{dbar0}) with $\Delta z=0.6 \langle z_g\rangle$.
\label{fig4}}
\end{figure}

In principle, one should use the modified fragmentation function evaluated 
according to the pQCD calculation for a dense medium. However, before
that can be done in a practical manner, we have used the effective
approach in Eq.~(\ref{dbar0}) by rescaling the fractional momentum by
$1-\Delta z$ to take into account of the parton energy loss. To verify
whether such an effective approach is adequate, we compare the
two modified fragmentation functions in Fig.~\ref{fig4}. We found
that the effective model (dashed lines) can reproduce the pQCD
result (solid lines) very well, but only when
$\Delta z$ is set to be $\Delta z\approx 0.6 \langle z_g\rangle$.
Therefore the actual averaged parton energy loss should be about
1.6 times of that used in the effective modified fragmentation function.
This difference is caused by the absorptive processes or unitarity
correction effect in the full pQCD calculation.

Unlike in DIS nuclear scattering, the dense medium in high-energy
heavy-ion collisions is not static. It has to go through rapid
expansion which should also affect the effective total parton energy
loss. The total energy loss extracted from experiments should be
a quantity that is averaged over the whole evolution history of the
expanding system. It is therefore useful to convert the averaged quantity
to an energy loss in a static system that has the same parton
density as the expanding system at its initial stage. If the
averaged total parton energy loss in a longitudinally expanding system 
with a transverse size $R$ is $\Delta E_{1{\rm d}}$, one finds \cite{gvw}
that the corresponding parton energy loss in a static system with the
same initial parton density would be 
$\Delta E=\Delta E_{1{\rm d}}(R/2\tau_0)$. Here $\tau_0$ is the initial
formation time of the dense medium.

\begin{figure}[t]
%\figurebox{20pc}{15pc}{} % to have a box alone
\epsfxsize=15pc 
% will enlarge or reduce the postscript figures based on the xsize
\centerline{\epsfbox{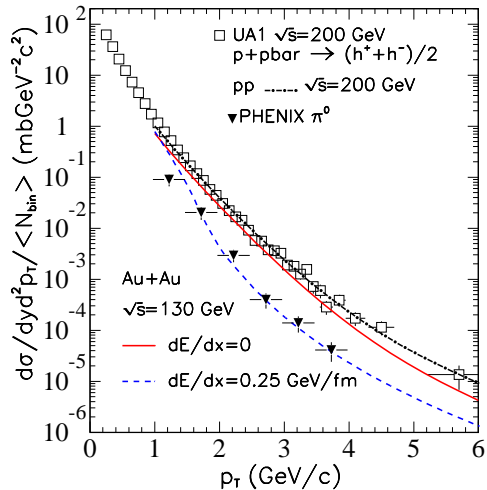}} % postscript image file name
\caption{pQCD parton model calculation of the charged hadron and 
pion spectra in $p\bar{p}$ and central $Au+Au$ collisions compared
with the experimental data \protect\cite{phenixpt,ua1}. The effective
modified fragmentation function is used in the calculation.
\label{fig5}}
\end{figure}

Comparing the recent PHENIX data \cite{phenixpt} with pQCD parton
model calculation
as shown in Fig.~\ref{fig5}, one can extract a value of $dE/dx$=0.25 GeV/fm
that one needs to use in the effective modified fragmentation function
in fitting the data. Taking into account the unitarity 
correction effect and the expansion, this
corresponds to an effective energy loss $dE/dx=1.6\times 0.25 R/2\tau_0$
in a static system with a density similar to the initial stage of the
expanding system at $\tau_0$. With $R\sim 6$ fm and $\tau_0\sim 1/p_0=0.2$ fm,
this would give $dE/dx\approx 12$ GeV/fm, which is about 40 times of
that in a cold nuclear matter as extracted from the DIS processes. Since
the parton energy loss is directly proportional to gluon density, this
implies that the gluon density in the initial stage of $Au+Au$
collisions is about 40 times higher than that inside a cold nucleus.

\section{Conclusions}

Recent RHIC data require a strong shadowing of gluon distribution
in nuclei within a two-component model of particle production in
heavy-ion collisions. Using this strong gluon shadowing with
an assumed impact-parameter dependence, the predicted centrality 
dependence of the hadron multiplicity agrees well with the recent
RHIC results. The results are compared with the parton
saturation model \cite{kn,kl}. In order to
differentiate the two models one needs more accurate experimental
data in both the most central and peripheral regions of centrality
or study the centrality dependence of the ratios at different
colliding energies.

It is argued that the jet quenching will not contribute much
to the total hadron multiplicity due to detailed balance in
a thermal environment, even though they significantly suppress
the high $p_T$ spectra.
The medium modification of the jet fragmentation
functions due to gluon radiation induced by the multiple parton
scattering can be calculated in pQCD. 
The predictions of the shape, energy dependence,
and the quadratic nuclear size $A^{2/3}$ dependence of
the modification agree well with the recent HERMES data. The
resultant parton energy loss in the cold nuclear medium is estimated
to be about 0.3 GeV/fm inside $Kr$ nuclei. Comparing to the
QCD result of the modification of fragmentation function, 
the actual averaged energy loss is found to be about 1.6 times that of the 
effective energy loss used in a earlier effective model for the same
modification. Considering the effect of expansion, the
recent PHENIX data imply a medium induced energy loss in central $Au+Au$
collisions equivalent to 12 GeV/fm in a static medium with the same
gluon density as in the initial stage of the collision.

\section*{Acknowledgments}
This work is supported by  the Director, Office of Energy 
Research, Office of High Energy and Nuclear Physics, 
Division of Nuclear Physics, and by the Office of Basic Energy Science, 
Division of Nuclear Science, of  the U.S. Department of Energy 
under Contract No. DE-AC03-76SF00098 and in part by 
NSFC under project 19928511 and 10075031.

\end{document}